\documentclass[reprint,superscriptaddress,amsmath,amssymb,aps,prl]{revtex4-2}
\usepackage{graphicx}% Include figure files
\usepackage{dcolumn}% Align table columns on decimal point
\usepackage{bm}% bold math
\usepackage{hyperref}% add hypertext capabilities
\usepackage{amsmath, amssymb, mathtools, isomath, nccmath}
\usepackage{siunitx}
\usepackage{xspace}
\usepackage{xcolor}
\usepackage[british]{babel}
\usepackage{sidecap}
\usepackage{soul}
\usepackage{multirow}

\newcommand{\Rb}{\ensuremath{^{87}\text{Rb}}\xspace}%
\usepackage{xcolor}

\sidecaptionvpos{figure}{h}

\begin{document}

% \title{\bf{Fast non-destructive cavity readout of single atoms within a coherent atom array}}
\title{\bf{Mid-circuit cavity measurement in a neutral atom array}}
\author{Emma Deist}
\thanks{These authors contributed equally to this work.}
\affiliation{Department of Physics, University of California, Berkeley, California 94720}
\affiliation{Challenge Institute for Quantum Computation, University of California, Berkeley, California 94720}
\author{Yue-Hui Lu}
\thanks{These authors contributed equally to this work.}
\affiliation{Department of Physics, University of California, Berkeley, California 94720}
\affiliation{Challenge Institute for Quantum Computation, University of California, Berkeley, California 94720}
\author{Jacquelyn Ho}
\affiliation{Department of Physics, University of California, Berkeley, California 94720}
\affiliation{Challenge Institute for Quantum Computation, University of California, Berkeley, California 94720}
\author{Mary Kate Pasha}
\affiliation{Department of Physics, University of California, Berkeley, California 94720}
\affiliation{Challenge Institute for Quantum Computation, University of California, Berkeley, California 94720}
\author{Johannes Zeiher}
\affiliation{Department of Physics, University of California, Berkeley, California 94720}
\affiliation{Max-Planck-Institut f\"{u}r Quantenoptik, 85748 Garching, Germany}
\affiliation{Munich Center for Quantum Science and Technology (MCQST), 80799 Munich, Germany}
\author{Zhenjie Yan}
\affiliation{Department of Physics, University of California, Berkeley, California 94720}
\affiliation{Challenge Institute for Quantum Computation, University of California, Berkeley, California 94720}
\author{Dan M. Stamper-Kurn}
\email[]{dmsk@berkeley.edu}
\affiliation{Department of Physics, University of California, Berkeley, California 94720}
\affiliation{Challenge Institute for Quantum Computation, University of California, Berkeley, California 94720}
\affiliation{Materials Sciences Division, Lawrence Berkeley National Laboratory, Berkeley, California 94720}
\date{\today}

% Word count:  main.tex was 2664 words.  Now: 2860/ 2820 / 2812 / 2719 / 2689 / 2642 
\begin{abstract}

Subsystem readout during a quantum process, or mid-circuit measurement, is crucial for error correction in quantum computation, simulation, and metrology. 
Ideal mid-circuit measurement should be faster than the decoherence of the system, high-fidelity, and nondestructive to the unmeasured qubits.
%A mid-circuit measurement must be fast and high-fidelity, and non-destructive to the unmeasured qubits.
% The non-destructive measurement of a subsystem within a larger quantum system is crucial for error correction during quantum computation, simulation, and metrology, and for studying open quantum system dynamics.  
%In many quantum technologies based on trapped atoms, measurement is performed by imaging all atoms simultaneously, a process that is typically slow and that decoheres the entire quantum system.  
Here, we use a strongly coupled optical cavity to read out the state of a single tweezer-trapped $^{87}$Rb atom within a small tweezer array.  Measuring either atomic fluorescence or the transmission of light through the cavity, we detect both the presence and the state of an atom in the tweezer, within only tens of microseconds, with state preparation and measurement infidelities of roughly $0.5\%$ and atom loss probabilities of around $1\%$.  Using a two-tweezer system, we find measurement on one atom within the cavity causes no observable hyperfine-state decoherence on a second atom located tens of microns from the cavity volume.
This high-fidelity mid-circuit readout method is a substantial step towards quantum error correction in neutral atom arrays.

% \blue{Using a two-atom array and simple circuit of a Ramsey sequence, we find that a mid-circuit measurement of one atom within the cavity causes no observable hyperfine-state decoherence on a second atom located tens of microns from the cavity volume.
%thus realizing the requirements for a mid-circuit measurement.
% This high-fidelity mid-circuit readout method is a substantial step towards fault-tolerant quantum computation with neutral atom arrays.
% } 
% \jh{This last sentence seems like a pretty bold claim. Maybe better to say something like ''Our high fidelity mid-circuit readout is an advancement towards fault-tolerant quantum computation with neutral atom arrays."}

\end{abstract}
\maketitle

%%%%%%%%%%%%%%%%%%%%%%%%%%%%%%%%%%%%%%%%%%%%%%%%%%%%%%%%%%%%%%%%%%%%%%%%%%%%
%                    Section: General Introduction                         %
%%%%%%%%%%%%%%%%%%%%%%%%%%%%%%%%%%%%%%%%%%%%%%%%%%%%%%%%%%%%%%%%%%%%%%%%%%%%

%General intro as to why our experiment is relevant, how it ties in with previous work, and in which ways it goes beyond previous work. How general this is will depend on the Journal.

Numerous applications of controlled many-body quantum systems require measurements that read out and affect only a part of the system, i.e.\ mid-circuit measurements. 
% Numerous applications of controlled many-body quantum systems require that part of the system be measured while other parts remain coherent. 
Examples include quantum error correction~\cite{Shor95qec,Stea96qec}, measurement-based quantum computation~\cite{Raus01oneway}, quantum-error-corrected metrology~\cite{Kessler2014,Dur2014,Zhou2020}, and an entanglement phase transition induced by mid-circuit measurements on a quantum circuit ~\cite{Li2018,Skinner2019}.
%\jh{Would it make sense to also mention quantum teleportation as an example?}
%\zy{let's define mid circuit measurement in first sentence}
Effective mid-circuit measurements should satisfy three requirements: They must be faster than the decoherence rate of the system, have low error rates (e.g.\ below around 1\% for implementing surface-code quantum error correction \cite{Dennis2002,Raussendorf2007,Fowler2012b}), and be sufficiently local so as not to disturb unmeasured quantum bits.

%\blue{Achieving  mid-circuit measurement is a challenge in several platforms for engineering large-scale quantum systems. The readout speed must be much faster than the decoherence rate of the system. The detection fidelity must be sufficient for realizing quantum feedback; for example implementing surface code quantum error correction requires an infidelity below $1\%$~\cite{Wang2011}. Lastly, the act of measurement must not destroy the coherence of the rest of system.} 

In atom-based systems such as atom-tweezer arrays~\cite{Barredo2016,Endres2016}, lattice-trapped atoms~\cite{Wang2015,Gros17qsimreview}, and trapped ion chains~\cite{Bruz19}, the many-atom state is often read out through optical fluorescence imaging. 
Practical limitations on the numerical aperture (NA) of imaging systems require many photons to be scattered by an atom before it is detected.  This requirement impairs the use of free-space imaging for mid-circuit measurement:  Measurements tend to be slow (e.g.\ on the order of 10~\cite{Fuhrmanek2011,Kwon2017,MartinezDorantes2017,Covey2019,Urech2022} or 100 ms~\cite{Gros21qgmdetection} in atomic tweezer arrays and quantum gas microscopes, respectively~\cite{FastImgFN}\nocite{Bergschneider2018,Xu2021}), of limited state-detection fidelity owing to spontaneous Raman transitions during detection; and destructive to nearby atoms that can absorb scattered photons.%.

\begin{figure}
    \centering
    \includegraphics{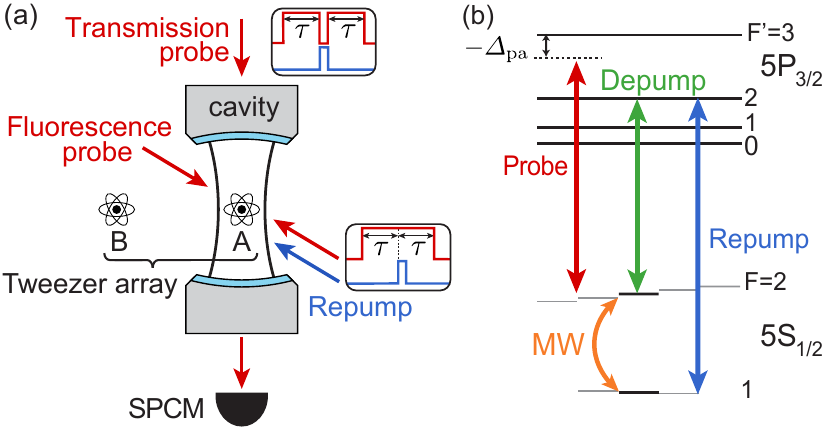}
    \caption{
    % \textbf{Overview of the experiment.}
    Experiment schematic.
    (a) Single atoms are loaded into each of two tweezers that can be translated perpendicularly to the cavity axis for individual readout.
    Counter-propagating fluorescence probe beams, and also a unidirectional repump beam, are focused on the atom inside the cavity mode. 
    The transmission probe beam couples directly into the cavity.
    % Both detection schemes use two consecutive probing intervals to first detect atoms in the bright $F=2$ manifold and then atoms in the $F=1$, by repumping them into the bright state before (transmission) or during (fluorescence) the second interval.
    (b)\Rb level structure. The probe beams (red) are detuned by $\Delta_\mathrm{pa}$ from the $F=2\rightarrow F'=3$ cycling transition. The repump and depump beams (blue and green) are on resonance with the $F=1\rightarrow F'=2$ and $F=2\rightarrow F'=2$ transitions. A resonant microwave pulse drives the  Zeeman-insensitive $|F=1, m_F = 0\rangle \rightarrow |F=2, m_F = 0\rangle$ hyperfine transition.
    }
    \label{fig:Fig_1}
\end{figure}

Here, we demonstrate mid-circuit optical detection of an atomic tweezer array wherein a single atom is measured with high fidelity while the remaining array retains quantum coherence.
%, realizing the requirements for a mid-circuit measurement.
For this, we use a strongly coupled cavity to detect a single optical tweezer, allowing for rapid, state-sensitive, high-fidelity, low-atom-loss local measurement with minimal photon scattering of about 100 photons.
% For this, we use a strongly coupled cavity to detect a single optical tweezer, allowing for rapid, state-sensitive, high-fidelity, low-atom-loss local measurement with minimal photon scattering of about 100 photons.
We benchmark our measurement with a two-atom tweezer array, measuring single atoms sequentially by translating each tweezer trap into the cavity mode, and then detecting light emitted by the cavity that is either fluoresced by the driven atom or transmitted through the driven cavity [Fig.\ \ref{fig:Fig_1}(a)].
We observe that an initially prepared hyperfine spin coherence of one atom persists even as the other atom is measured at high fidelity. %\eg{Maybe this last sentence could be tweaked or maybe that would be overdoing it with "mid-circuit" everywhere.}

%%%%%%%%%%%%%%%%%%%%%%%%%%%%%%%%%%%%%%%%%%%%%%%%%%%%%%%%%%%%%%%%%%%%%%%%%%%%
%       Figure 1: Schematic explaining platform and measurement            %
%%%%%%%%%%%%%%%%%%%%%%%%%%%%%%%%%%%%%%%%%%%%%%%%%%%%%%%%%%%%%%%%%%%%%%%%%%%%
%Note: Make figures directly with correct dimensions (check Journal for dimensions. There are 1 col figs, 1.5 col figs or 2 col figs. Do NOT scale in tex, as that makes all the text in the figures differently sized. Generally, figure labels like "a" or "b" should be 8-9pt, axis labels min 6pt, better 8pt. Font should be HeleveticaNeue. Datapoints look nice with a brighter inside and a darker edge. The errorbars should have the same color as the edge. All ticks, lines and errorbars should be rounded. Solid theory predictions or fits should have the color of the errorbars.

Our experimental setup is described in Ref.\ \cite{Deist2022}.
Briefly, a bulk optically trapped gas of ultracold $^{87}$Rb atoms is prepared near the volume of a horizontal-axis, near-concentric in-vacuum Fabry-P\'{e}rot optical cavity with a mirror spacing of 9.4 mm.
Atoms are loaded into optical tweezer traps formed by 808-nm-wavelength light that is projected vertically through a high-NA imaging system.
An acousto-optical deflector (AOD) allows us to generate multiple traps in a one-dimensional array and to translate them perpendicularly to the cavity axis.
We illuminate the tweezers with counter-propagating light that is detuned about $2\pi\times 35$ MHz below the $D_2$ $F=2 \rightarrow F^\prime = 3$ laser-cooling transition, and also with repump light, resonant with the $F=1 \rightarrow F^\prime = 2$ transition, both at a wavelength of 780 nm.
This illumination reduces the population in each tweezer to either zero (empty tweezer) or one atom, which we distinguish by imaging the resulting fluorescence through the high-NA objective.

Single tweezer-trapped atoms can serve as long-lived qubits by encoding quantum information in the ground-state hyperfine spin~\cite{Xia2015,Levine2019}.
Following this approach, we prepare our atoms into the $F=1$ or $F=2$ manifold by applying either depump ($F=2 \rightarrow F^\prime =2$) or repump light, respectively~\cite{SI}\nocite{Schlosser2001,Tuchendler2008,Kaufman2012,Gerb20thesis,Qutip} [Fig.~\ref{fig:Fig_1}(b)].
Combined with information from the aforementioned fluorescence image, the tweezers are thereby prepared in one of three tweezer states: empty, containing an atom in the $F=1$ manifold, or containing an atom in the $F=2$ manifold. % As quantified below, this state preparation is nearly perfect.
We use our cavity to measure a single tweezer, distinguishing each of these three tweezer states.
The cavity reaches the single-atom strong coupling regime, with a cooperativity on the $^{87}$Rb $D_2$ cycling transition of $C = g_0^2/(2 \kappa \gamma) = 2.3$ with $\{g_0, \kappa, \gamma\} = 2\pi \times \{2.7, 0.53, 3.0\}$ MHz.
Here, $g_0$ is the maximum atom-photon coupling strength between the $F=2$ and $F^\prime=3$ stretched states at a field antinode in the center of the TEM$_{00}$ cavity mode with a beam waist of $w_0 = 20(3)\,\mu\textrm{m}$.
The half-linewidths of the cavity and atomic resonances are $\kappa$ and $\gamma$, respectively.
%\blue{Both $g_0$ and $\kappa$ are much smaller than the ground and excited state hyperfine splittings, leading to a low rate of $F=2 \rightarrow F = 1$ spontaneous Raman transitions; for comparison, see Ref.\ \cite{Dordevic2021}.}

Our high-cooperativity cavity supports two measurement methods.
In the fluorescence method, we directly illuminate the atom and collect its fluorescence using the cavity.
Strong atom-cavity coupling results in a large collection efficiency into a single optical mode that is detected with little background noise.
In the transmission method, we drive the cavity near its resonance and measure the transmission of cavity probe light.
Here, atom-cavity hybridization causes a single atom to broaden (at low $C$) or split (at high $C$) the cavity resonance line, reducing the transmitted intensity.
Single-atom detection using strongly coupled cavities has been demonstrated previously, both through fluorescence~\cite{Bochmann2010,Gallego2018} and cavity transmission or reflection~\cite{Bochmann2010, Gehr2010}.
For a two-atom array, collective detection and one-way transport from a cavity into free space has been demonstrated in Ref.~\cite{Dordevic2021}, while probabilistic atom-photon conversions with single-atom addressability has been shown in Ref.~\cite{Langenfeld2020}.
% Cavity detection has also been demonstrated with a two-atom tweezer array capable of one-way transport from a photonic crystal cavity to a free space array region~\cite{Dordevic2021}.
%The present work extends these results to measurements on atoms trapped and transported in fully configurable optical tweezers, realizing a mid-circuit cavity measurement of a single atom that does not decohere the rest of the array, as well as advancing the state-of-the-art with faster and higher-fidelity state detection.
The present work extends these results to high-fidelity single-atom state detection that does not decohere the rest of the array, demonstrating the necessary features of a mid-circuit measurement in a neutral atom quantum information processor. 
% \LL{I think we should also mention the fundamental readout speed limit that we have obtained as a push through in this work.}
% \jh{I felt like ''mid-circuit measurement" was kind of lost in the middle of the old sentence, so I expanded it to make it more obvious. Might be a bit wordy now, though.}

\begin{figure}
    \centering
    \includegraphics{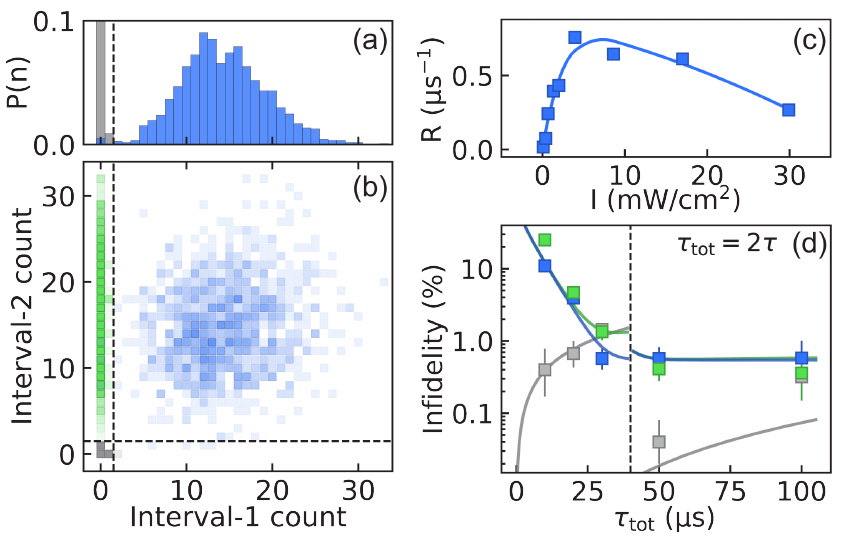}
    \caption{Fluorescence measurement. A single-probe histogram (a) and two-probe scatter plot (b) show the detected photon counts for tweezers in the no-atom (gray), $F=2$ atom (blue), or $F=1$ atom (green) state, taken with $\tau = 25\, \mu$s and $\Delta_\mathrm{pc} = -2 \pi \times 10$ MHz.  The threshold (dashed line) between high and low fluorescence is set between 1 and 2 detected photons.  (c) The optimal probe intensity $I$ yields a maximum high count rate of $R = 0.76\,\mu\mathrm{s}^{-1}$. Solid line is a guide to the eye. (d) SPAM infidelity is determined for total measurement times $\tau_\mathrm{tot}=2\tau$ from $10$ to $100\, \mu$s.
    % Data for $\tau_\mathrm{tot} \leq 30\, \mu$s use a lower detection threshold (between 0 and 1 detected photons).
    Solid lines are fits to a model described in~\cite{SI}.
    For $\tau_\mathrm{tot} \leq 40\, \mu$s, indicated by the vertical gray line, both the data and model are calculated using a lower detection threshold between 0 and 1 photons. 
    % \textbf{(a) Photon count histogram} of with/without  atom fluorescence into cavity (blue/gray).
    % \textbf{(b) 2D scatter plot}  of two interval photon counts. The dashed lines at low-high threshold of 1.5 divide all of the points into four sections,  bench-marking the state detection infidelity for each case (see Table~\ref{tab:table1} for details). 
    % \textbf{(c) Photon detection rate} vs fluorescence probe intensity.
    % \textbf{(d) Measurement infidelity} at various probe interval times $\tau$ under optimal probe Rabi frequency of $XXX$, reaches $99.XX$ at around $\tau=XX$. The parameters for the above plots are: $\Delta_{PC}=-10\,\mathrm{MHz}$, $\tau=25\,\mu \mathrm{s}$ (a-b), $I=XX$(d).
    }
    \label{fig:Fig_F}
\end{figure}

\begin{figure}
    \centering
    \includegraphics{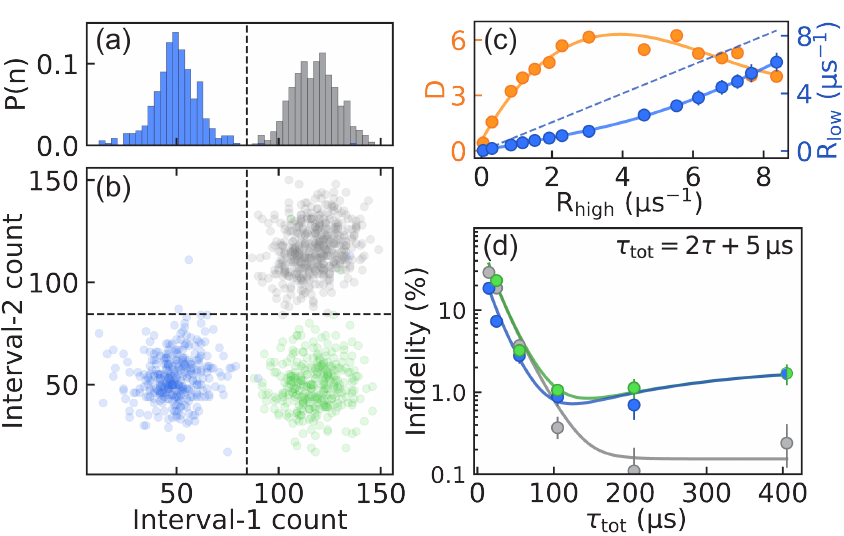}
    \caption{
    Transmission measurement.
    A single-probe histogram (a) and two-probe scatter plot (b) show the detected photon counts for tweezers in the no-atom (gray), $F=2$ atom (blue), or $F=1$ atom (green) state, taken with $\tau = 50\, \mu$s and $\Delta_\mathrm{pc} = 0$.
    The threshold (dashed line) between high and low transmission is set above 77 detected photons.  (c) The transmitted photon count rate with an $F=2$ atom in the cavity ($R_\mathrm{low}$, blue) is lower than the rate ($R_\mathrm{high}$, x axis and dotted line) observed without.
    Ashman's $D$ (orange), a measure of the separation between $R_\mathrm{low}$ and $R_\mathrm{high}$, reaches a maximum owing to atomic saturation.
    Lines are guides to the eye.
    (d) SPAM infidelity is determined for each of the initial tweezer states, with total measurement times $\tau_\mathrm{tot} = 2 \tau + 5 \, \mu$s ranging from 15 to 205 $\mu$s.
    The threshold between high and low is selected to minimize infidelity at each $\tau$. Lines are fits to a model described in~\cite{SI}.
}
    \label{fig:Fig_T}
\end{figure}

%%%%%%%%%%%%%%%%%%%%%%%%
% Two window scheme %
%%%%%%%%%%%%%%%%%%%%%%%%
In both measurement methods, our goal is to realize three-state sensitivity with measurement infidelity at the subpercent level, as required in certain protocols for quantum error correction~\cite{Dennis2002,Raussendorf2007,Fowler2012b}. We do this by probing the atom-cavity system in two consecutive probe intervals.
In each interval of duration $\tau$, using probe light near the $F=2 \rightarrow F^\prime=3$ transition, we determine whether the cavity contains a single atom in the $F=2$ manifold.
This is done by counting photons emitted from the cavity using a single-photon counting module (SPCM) with a total quantum efficiency of $\eta = 0.25$~\cite{SI}.
The detection path and SPCM are polarization-insensitive.
A positive detection of an $F=2$ atom is indicated by the observed photon number being either higher [fluorescence, see Fig.\ \ref{fig:Fig_F}(a)] or lower [transmission, see Fig.\ \ref{fig:Fig_T}(a)] than an optimized threshold.
The second probe interval begins with (for fluorescence) or is preceded by (for transmission) a  $\tau_{\mathrm{rp}}=5\,\mu\mathrm{s}$ pulse of localized repump light % local illumination of the detected tweezer with repump light for $\tau_{\mathrm{rp}}=5\,\mu\mathrm{s}$, driving an atom in the $F=1$ manifold to the $F=2$ manifold
[Fig.~\ref{fig:Fig_1}(a) insets]. 
% Before (for transmission) or beginning with (fluorescence) the second probe interval, we drive the detected atom from the $F=1$ to the $F=2$ manifold by illuminating the tweezer locally with repump light for $\tau_{\mathrm{rp}}=5\,\mu \mathrm{s}$.
The negative detection of an $F=2$ atom in the first interval followed by a positive detection in the second interval measures the tweezer as having contained an $F=1$ atom, whereas a negative detection in both intervals measures the tweezer as being empty.

%%%%%%%%%%%%%%%%%%%%%%%%
% Fluorescence section %
%%%%%%%%%%%%%%%%%%%%%%%%

%%%%%%%%%%%%%%%%%%%%%%%%%%%%%%%%%%%%%%%%%%%%%%%%%%%%%%%%%%%%%%%%%%%%%%%%%%%%
%               Section: Figs 2,3; Table                                   %
%%%%%%%%%%%%%%%%%%%%%%%%%%%%%%%%%%%%%%%%%%%%%%%%%%%%%%%%%%%%%%%%%%%%%%%%%%%

In the cavity fluorescence method, we set the cavity resonance frequency $\omega_\mathrm{c}$ to be detuned by $\Delta_\mathrm{ca} = \omega_\mathrm{c}- \omega_\mathrm{a} = - 2 \pi \times 10$ MHz below the laser-cooling transition frequency $\omega_a$.
We illuminate the atom with vertically counter-propagating probe beams in a lin-perp-lin configuration in order to provide polarization gradient cooling during measurements.
% , focused to beam waists of roughly 3 and 13 $\mu$m, in a lin-perp-lin configuration.
The probe frequency $\omega_\mathrm{p}$ is tuned slightly below the cavity resonance ($\Delta_\mathrm{pc} = \omega_\mathrm{p} - \omega_\mathrm{c} \sim -\kappa/2$) to realize cavity cooling of the atomic motion ~\cite{Nuss053d}.
The probe light intensity is set to maximize the photodetection rate $R=R_\mathrm{max}$ of an $F=2$ tweezed atom in the cavity [Fig.\ \ref{fig:Fig_F}(c)]; lower probe intensity drives the atom below saturation, whereas higher probe intensity shifts the incoherent fluorescence spectrum outside the bandwidth of the cavity~\cite{Mollow1969,Kimble1976}.  
Experimentally, we find $R_\mathrm{max}\simeq 0.76\,\mu\mathrm{s}^{-1}$, which is below the theoretical maximum of $R_0 = \eta g_0^2/(4 \kappa) = 5.4\,\mu\mathrm{s}^{-1}$ predicted for a two-level atom~\cite{SI}.
This difference may be explained by two effects.
First, the tweezer-trapped atom is poorly localized along the cavity axis, exhibiting rms position fluctuations of up to 200 nm with respect to the standing-wave pattern (periodicity of 390 nm) of the cavity mode; see Ref.\ \cite{Deist2022}.  
The effective square of the atom-cavity coupling strength is thus averaged roughly to $g_\mathrm{eff}^2 \simeq g_0^2/2$ owing to spatial random sampling.
Second, internal state dynamics induced by the probe light drives the atom between Zeeman sublevels of the ground and excited states, reducing the effective time-averaged coupling to the two polarization modes supported by the cavity.
We estimate this effect reduces the maximum cavity emission rate by an additional factor of 0.28~\cite{SI}.

Fluorescence measurement outcomes, obtained after preparing a single intracavity tweezer in each of the three tweezer states, are shown in Fig.\ \ref{fig:Fig_F}.
For a probe interval of $\tau= 25\,\mu$s, we observe a large contrast between the photon number detected for a tweezer prepared in the $F=2$ state, and that detected for either the no-atom or $F=1$ states [Fig. \ref{fig:Fig_F}(a)].
Combining data from two consecutive 25 $\mu$s probe intervals (total measurement time of $\tau_\mathrm{tot} = 2 \tau = 50\,\mu$s), and setting the threshold for state detection between 1 and 2 photons, we achieve a state preparation and measurement (SPAM) error of several times $10^{-3}$ for each of the three initial tweezer states (Table \ref{tab:table1}).
For shorter $\tau$ [Fig.\ \ref{fig:Fig_F}(d)], statistical fluctuations in the detected photon number lead us to misidentify bright states as dark states in either the first or second probe intervals, leading to infidelity in $F=2$ and $F=1$ state detection, respectively.
For longer $\tau$, state preparation error and false detection error caused by the depumping of an $F=2$ atom before detecting an above-threshold number of photons set a limit on the achievable fidelity.
We estimate that these two error sources contribute roughly equally to the overall SPAM error~\cite{SI}.
Table \ref{tab:table1} also reports low atom loss probabilities on the order of $1\%$, with higher loss rates for atoms in the $F=2$ manifold due to scattering-induced heating through both probe intervals. 
% One expects the latter error to increase for probe light increasingly detuned from the atomic resonance; fluorescence measurements made with large $|\Delta_\mathrm{ca}|$ support this expectation~\cite{SI}.\eg{?}

\begin{table}[b]%The best place to locate the table environment is directly after its first reference in text
\caption{\label{tab:table1}%
Measurement infidelity and loss probability}
\small
\begin{tabular}{|l|l|c|c|c|}
\hline
\multicolumn{2}{|c|}{} & No atom & F=1 & F=2\\
\hline
\multirow{3}{2cm}{Fluorescence $2\times(\tau=25\,\mu\mathrm{s})$}&
\textrm{Outcome}&
\textrm{low-low}&
\textrm{low-high}&
\textrm{high-X}\\
\cline{2-5}
\newline & \textrm{Infidelity} & 0.04(3)\,\% & 0.4(2)\,\% & 0.6(2)\,\% \\
\cline{2-5}
\newline  & \textrm{Loss prob.} &  n\slash a & 0.2(2)\,\% & 1.4(3)\,\% \\
\hline
\multirow{3}{2cm}{Transmission $2\times(\tau=50\,\mu\mathrm{s})\newline+5\,\mu \mathrm{s}$}&
\textrm{Outcome}&
\textrm{high-high}&
\textrm{high-low}&
\textrm{low-X}\\
\cline{2-5}
\newline  & \textrm{Infidelity} & 0.4(1)\,\% & 1.1(2)\,\% & 0.9(2)\,\% \\
\cline{2-5}
\newline  & \textrm{Loss prob.} & n\slash a & 0.7(3)\,\% & 1.4(2)\,\% \\
\hline
\end{tabular}
\end{table}

%%%%%%%%%%%%%%%%%%%%%%%%
% Transmission section %
%%%%%%%%%%%%%%%%%%%%%%%%

In the cavity transmission method, we drive the cavity with light that is resonant with both the cavity and the atom ($\Delta_\mathrm{ca} = \Delta_\mathrm{pc}=0$).
The circularly polarized probe light, together with a weak magnetic field applied along the cavity axis, pumps $F=2$ atoms into the spin-stretched state, maximizing their coupling to the cavity.
For weak probe light, we observe that an $F=2$ atom in the cavity reduces the detected transmitted photon rate $R_\mathrm{low}$ to 0.4 times the rate $R_\mathrm{high}$ observed with either no atoms or an $F=1$ atom in the cavity.
For low saturation, one would expect $R_\mathrm{low}/R_\mathrm{high} = (1+2 C)^{-2}$ for fixed atom-cavity coupling strength. Averaging this expression over a uniform atomic spatial distribution along the cavity axis yields $R_\mathrm{low}/R_\mathrm{high} = 0.27$ for our system.
% For constant atom-cavity coupling strength, one expects $R_\mathrm{low}/R_\mathrm{high} = (1+2 C)^{-2}$ in the low saturation regime. Averaging for a uniform spatial distribution along the cavity axis yields $R_\mathrm{low}/R_\mathrm{high} = 0.27$ for our system. 
The difference between the observed and expected transmission reduction may be explained by an inhomogeneous broadening of the atomic resonance of roughly 4 MHz, caused by the ac Stark shift of the tweezer trap light~\cite{Bochmann2010}.  
At high probe intensity, atomic saturation leads to $R_\mathrm{high} - R_\mathrm{low}$ reaching a constant difference of roughly $2.4 \,\,\mu\mathrm{s}^{-1}$. At an intermediate probe intensity setting of $R_\mathrm{high}\simeq 2.2\,\,\mu\mathrm{s}^{-1}$, the bimodal separation statistic $D$~\cite{AshmanD}\nocite{Ashman} between the high and low photon count distributions reaches its maximum [Fig.\ \ref{fig:Fig_T}(c)]. %\zy{I am sure now this is not what we see in experiment, as we are seeing a strong super-Poissonian distribution for with atom detection. Let's just stick with experimental results.}.\LL{Alternative try below}
% \LL{The difference between the observed and expected transmission reduction may be explained by an inhomogeneous broadening on the atomic resonance of roughly 4 MHz, caused by the ac Stark shift of the tweezer trap light~\cite{Bochmann2010}.  
% At high probe intensity, atomic saturation leads to $R_\mathrm{high} - R_\mathrm{low}$ reaching a maximum difference of $\eta\gamma/2C$ for constant cooperativity. The detection signal to noise ratio (SNR), defined as the difference of mean divided by the sum of standard deviation, reaches a maximum at an intermediate setting of $R_\mathrm{high}\simeq 2.2\,\,\mu\mathrm{s}^{-1}$ [Fig.\ \ref{fig:Fig_T}(b)]}

Transmission measurements made at this optimal probe intensity, with two probe intervals of $\tau=50\, \mu$s each ($\tau_\mathrm{tot}
= 2 \tau + 5\,\mu\mbox{s}=105\,\mu$s), again show clear distinctions among tweezers prepared initially in each of the three tweezer states [Fig.\ \ref{fig:Fig_T}(a)].
The detection infidelities and atom loss (Table\ \ref{tab:table1}) are comparable to those obtained through fluorescence.
However, the smaller contrast between high and low detection rates causes the transmission method to be generically slower than the fluorescence method of detection.
Transmission measurements with a higher $C$ would be interaction-free~\cite{Kwiat1995}, thus suppressing depumping errors and mechanical effects from light scattering, which provides particular advantages for detecting trapped particles, such as single molecules~\cite{Anderegg2019,Zhang2021}, that lack a cycling optical transition.

%eliminating the possibility that the atoms in the coupled state could undergo depumping during measurement. This makes the cavity transmission method a good candidate for detecting trapped particles without a good cycling transition, such as a single molecule or transition metal atom~\cite{Eustice2020a} in an optical tweezer, and would guarantee a known final state of the detected particle for reuse. \dmsk{I removed the reference to the transition metals.  In fact, these have prefectly good ($10^6$ photons) cycling transitions}  

%%%%%%%%%%%%%%%%%%%%%%%%%%%%%%%%%%%%%%%%%%%%%%%%%%%%%%%%%%%%%%%%%%%%%%%%%%%%
%                         Section: Fig 4                                   %
%%%%%%%%%%%%%%%%%%%%%%%%%%%%%%%%%%%%%%%%%%%%%%%%%%%%%%%%%%%%%%%%%%%%%%%%%%%%

\begin{figure}
    \includegraphics{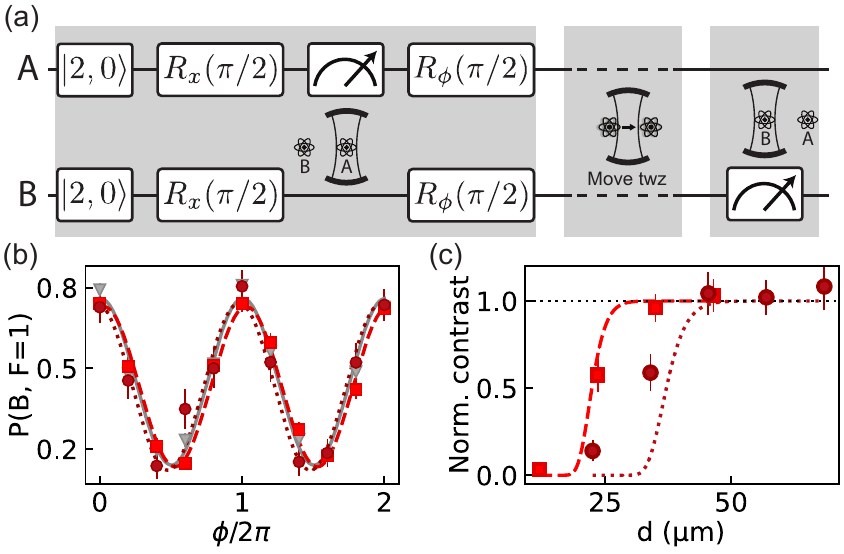} 
    \caption{Mid-circuit measurement. (a) Quantum circuit representing a Ramsey sequence with a mid-circuit measurement of atom A. Atom A (B) is initially located inside (outside) the cavity. Both atoms undergo two $\pi/2$ rotation pulses with variable relative phase $\phi$, denoted as $R_x(\pi/2)$ and $R_\phi(\pi/2)$. Atom A is measured between the two pulses using either fluorescence or transmission measurement methods.
    % \st{Both are prepared in a two-hyperfine-state superposition using a microwave $\pi/2$ pulse.  Atom A is measured with a total measurement duration $\tau_\mathrm{tot}$.}
    Then, both tweezers are repositioned before atom B is measured using the same method.  (b) The $F=1$ state probability of atom B, $P({\mathrm{B}, F=1})$, shows Ramsey fringes as the phase $\phi$ of the second pulse is varied. We observe no distinction between Ramsey fringes measured following fluorescence (light red squares, $\tau = 25 \, \mu$s) and transmission (dark red circles, $\tau = 50 \, \mu$s) detections of atom A, and no detection of atom A (gray). Normalized contrast is defined as the ratio of the Ramsey fringe contrasts observed with and without a mid-circuit measurement on atom A.  (c) Normalized contrast vs.\ the initial distance of atom B from the cavity center, with $\tau$ settings as in (b). The dashed (dotted) line is a theoretical estimate based on the intensity and size of the fluorescence probe beam (cavity mode)~\cite{SI}.
   }
\label{fig:Fig_4}
\end{figure}

% Next, we demonstrate that our cavity-enhanced detection is not only fast and high-fidelity, but also local, a third requirement for a mid-circuit measurement.

Next, we demonstrate that our cavity-enhanced detection of one atom does not perturb the quantum evolution of other atoms in an array, an essential requirement for a mid-circuit measurement. 
%\jh{I like Emma's version of this sentence more. We are building up an argument for why our measurement should be considered mid-circuit, so it doesn't make sense to suddenly call it that.}
We implement a simple quantum circuit consisting of single-qubit gates, realizing a Ramsey sequence on a two-atom tweezer system [Fig.~\ref{fig:Fig_4}(a)].  We form the array with atom A initially trapped within, and atom B at a variable radial distance $d$ outside, the cavity. 
Both atoms are initialized in the $|F=2, m_F=0\rangle$ state and subject to a $\pi/2$ microwave-induced rotation to the $|F=1, m_F = 0\rangle$ state~\cite{SI}.
A mid-circuit measurement is performed on atom A, using either detection method with the optimal probe times in Table \ref{tab:table1}.
%The circuit is then completed 
We complete the circuit by applying a second $\pi/2$ pulse with a variable phase offset $\phi$, translating atom B into and atom A out of the cavity simultaneously within 200 $\mu$s, and performing a cavity measurement of atom B.
% \jh{just trying to avoid saying ''complete" twice!}

Measurements on atom B show a characteristic Ramsey fringe as $\phi$ is varied [Fig.\ \ref{fig:Fig_4}(b)].
We quantify the effect of mid-circuit measurement by considering a normalized contrast, taken as the ratio of the Ramsey-fringe contrasts with and without mid-circuit measurement \footnote{The limited fringe contrast of about 0.6 observed in our setup even without mid-circuit detection arises from imperfect preparation in the $m_F=0$ magnetic sublevel and ac Stark shifts of the microwave frequency by tweezer light.}. 
We observe a normalized Ramsey contrast above $97\%$ with $84\%$ confidence level~\cite{SI}, when atom B is $d=34.5\,\mu\mathrm{m}$ ($d=46.0\,\mu\mathrm{m}$) away from the cavity mode center for fluorescence (transmission) measurement.
A mid-circuit fluorescence measurement begins to affect the coherence of atom B once atom B is within about $20\,\mu\mathrm{m}$ of the cavity center~\cite{CrainFN}\nocite{Crain2019}.
This length scale is consistent with the beam waists of the fluorescence probe beams~\cite{SI}.
A transmission measurement begins to affect atom B at a larger distance of roughly $35\,\mu\mathrm{m}$ from the cavity center, consistent with the beam waist of the cavity mode~\cite{Deist2022}.

% avoid over use of mid-circuit detection

%\blue{To quantify additional loss of coherence caused by mid-circuit measurement, we determine a normalized contrast, taking the ratio of the fringe contrasts observed with and without an intermediate measurement of atom A.} %\eg{something using specific language of ``circuit result'' (or whatever the right term is) with and without the mid-circuit measurement?} \jh{again, I have some qualms about calling the measurement mid-circuit in this paragraph...}

% %%%%%%%%%%%%%%%%%%%%%%%%%%%%%%%%%%%%%%%%%%%%%%%%%%%%%%%%%%%%%%%%%%%%%%%%%%%%
% %                   Section: Conclusion and outlook                        %
% %%%%%%%%%%%%%%%%%%%%%%%%%%%%%%%%%%%%%%%%%%%%%%%%%%%%%%%%%%%%%%%%%%%%%%%%%%%%

% \eg{Tried to recast in terms of circuit stuff which meant moving stuff around and taking out general comments about cavity measurement for atom/molecule tweezer platform..
% But the point about integration of cavity measurement with atom/molecule tweezer platform is still true.. it's easier to imaging putting our cavity around a tweezed molecule platform than the PCC stuff, at least me. I think our experimental platform is still noteworthily different from the 2-tweezer Lukin work to comment on, even though it's not our main point. thoughts?}
Our work demonstrates that the integration of cavity-enhanced measurement with a configurable tweezer array enables mid-circuit measurement within a neutral atom quantum information platform. 
%\eg{are we totally sure claiming this?} 
We achieve measurement infidelities comparable to the best previous results in atomic tweezer systems~\cite{Covey2019}, in a manner that not only allows subsystem-selective measurement but is also fast, with the measurement time being shorter than not only the second-scale hyperfine-state coherence of tweezer-trapped atoms~\cite{Bluvstein2021b,Norcia2019}, but also the $\sim 100 \, \mu$s lifetime of the Rydberg states commonly used in Rydberg-tweezer systems~\cite{Beterov2009}.
Combined with the low probability of losing a trapped atom during detection, cavity-based measurement could also enable the deterministic preparation of atom arrays assembled atom-by-atom, without requiring free-space imaging and re-sorting~\cite{Miro2006,Fortier2007,Kim16array,Barredo2016,Endres2016}.
%\eg{this is a number pulled out from the response to referee... probably needs a bit more specificity?} \jh{this reference considers total gate error thresholds, so I wonder if we need to flush out the connection to measurement error rates? I think they include measurement error in their error model, but I have to read more carefully to figure out what the actual percentage for measurement error is...}

% \eg{do we still want any more general comment about atom/molecule tweezer platforms and using this for atom array preparation? Maybe we just cut that since it is not in the new focus}
%\dmsk{I don't really understand what we're saying in this last phrase.  I don't see the harm in just removing it}, and naturally integrates in the recently-presented quantum computing platform based on moving tweezer arrays~\cite{Bluvstein2021b}.

The detection time, infidelity, and loss of our measurement could be reduced further by several experimental improvements.
%Increasing $C$ would increase the maximum steady-state rate of fluorescence emission into the cavity as well as the contrast between transmission measurement levels, enabling faster detection. However, when the detection timescale approaches $1/\kappa$, our steady-state treatment does not apply. 
Increasing $g_0$ and $\kappa$ simultaneously, up until the onset of hyperfine-state mixing \cite{Dordevic2021}, would allow for more efficient and faster detection of scattered photons.
%, achieving higher $C$ and faster buildup and emission of cavity photons, would allow for continued improvement of detection speed. 
Better constraints on atomic motion, achieved by improved laser cooling~\cite{SI} or by stronger confinement along the cavity axis, would mitigate the effective motional reduction of atom-cavity coupling that we presently observe.
Speed limits imposed by the need to transport atoms into the cavity prior to measurement could be improved by employing optical-lattice-based conveyors~\cite{Kuhr2001}.
Transport could be eliminated altogether by maintaining the tweezer array entirely within the cavity volume and using rapid ac Stark shifts realized with local illumination to bring atoms selectively into resonance with the cavity for detection~\cite{Urech2022}. 
% \LL{Can we combine the previous two sentences? Feel like that can shorten it a bit. Here's a try;}
% \LL{Speed limits imposed by the need to transport atoms into the cavity prior to measurement could be improved/eliminated by employing optical-lattice-based conveyors~\cite{Kuhr2001}, or using local addressing beams to bring atoms selectively into resonance with the cavity for detection~\cite{Urech2022}.}
%\eg{Seems like this becomes especially important for mid-circuit measurement.. transport time can't be ignored. not sure if we need to change the wording here or if it's fine as is.}
% The atom loss probability, higher for atoms in the $F=2$ manifold due to scattering-induced heating during both probe intervals, could be reduced to The atom loss probability, higher for atoms in the $F=2$ manifold due to scattering-induced heating during both probe intervals, could be reduced at least to the subpercent level of atoms in the $F=1$ manifold using an adaptive measurement that stops each probe interval as soon as a measurement outcome is obtained. the subpercent level of atoms in the $F=1$ manifold using an adaptive measurement that eliminates the second probe interval for atoms already measured as $F=2$.
%The atom loss probability, higher for atoms in the $F=2$ manifold, could be reduced at least to the subpercent level of atoms in the $F=1$ manifold using real-time processing and an adaptive measurement that stops each probe interval when a measurement outcome is obtained. 
The atom loss probability could be reduced by using real-time processing and an adaptive measurement that stops each probe interval when a measurement outcome is obtained \cite{Chow2022}, and also by applying laser cooling briefly after detection. 

% \eg{We haven't really shown anything about local laser cooling so wouldn't this not really work for mid-circuit? maybe cut the second clause?}
% \\blue{This paragraph is all about improving the single-atom measurement.. anything to say about the ``mid-circuit'' element here? Maybe just something about how if the single atom measurement is faster then you can fit more in your circuit and the decoherence to the rest of the array will be less because you don't turn on the probe light for as long?} 
%as soon as an atom in the $F=2$ manifold is detected. 
%\eg{I know this is a little awkward; simplest idea is to not do the second probe window for F=2 atoms; better idea is to stop each probe window as soon as enough photons have been scattered - so time of measurement would be varied, but scattered photons would be minimized.}

\begin{acknowledgements}
We thank C.\ Liu for assistance in the lab and J.\ Gerber for comments on the manuscript. 
We acknowledge support from the AFOSR (Grant No. FA9550-19-10328), from ARO through the MURI program (Grant No. W911NF-20-1-0136), from DARPA (Grant No. W911NF2010090), and from the NSF QLCI program through grant number OMA-2016245. E.D.\ acknowledges support from the NSF Graduate Research Fellowship Program. J.H.\ acknowledges support from the NIH Molecular Biophysics Training Grant (Grant No. 5T32GM008295-31).
J.Z.\ acknowledges support from the BMBF through the program ``Quantum technologies - from basic research to market" (Grant No. 13N16265).

\end{acknowledgements}

% \bibliography{cavity-measurement,allrefs_x2}

\bibliography{cavity-measurement}

\begin{thebibliography}{58}
\bibitem [{Ash()}]{AshmanD}%
  \BibitemOpen
  \href@noop {} {}\bibinfo {note} {Given two distributions with means
  $\mu_{1,2}$ and standard deviations $\sigma_{1,2}$, Ashman's $D$ statistic
  for bimodal separation is defined as $D = |\mu_1-\mu_2|/\sqrt{(\sigma_1^2 +
  \sigma_2^2)/2}$; see Ref.\ \cite{Ashman}.}\BibitemShut {Stop}%
\end{thebibliography}

\clearpage

\end{document}